\documentclass{article}
\usepackage{amsmath}
\usepackage{amsfonts}

\usepackage[space]{cite}
\def\@biblabel#1{[#1]}

\usepackage{hyperref}
\usepackage{graphicx}
\usepackage[utf8]{inputenc}

\title{A Passion for Theoretical Physics: a special issue in memory 
of Peter G O Freund}

\author{Jeffrey A Harvey\thanks{Enrico Fermi Institute and Department of Physics, 
University of Chicago, 933 E 56th St., Chicago, IL 60637 USA}, 
Emil J Martinec${}^{*}$
and Rafael I Nepomechie\thanks{Physics Department, P.O. Box 248046, University of Miami, 
Coral Gables, FL 33124 USA}}

\begin{document}

\maketitle

\begin{abstract}
This is a preface to A Passion for Theoretical Physics, a special issue collection
of articles published in J. Phys. A in memory of Peter G O Freund.
\end{abstract}

\setcounter{footnote}{0}

\begin{figure}[ht]
\begin{center}
\includegraphics[scale=0.1]{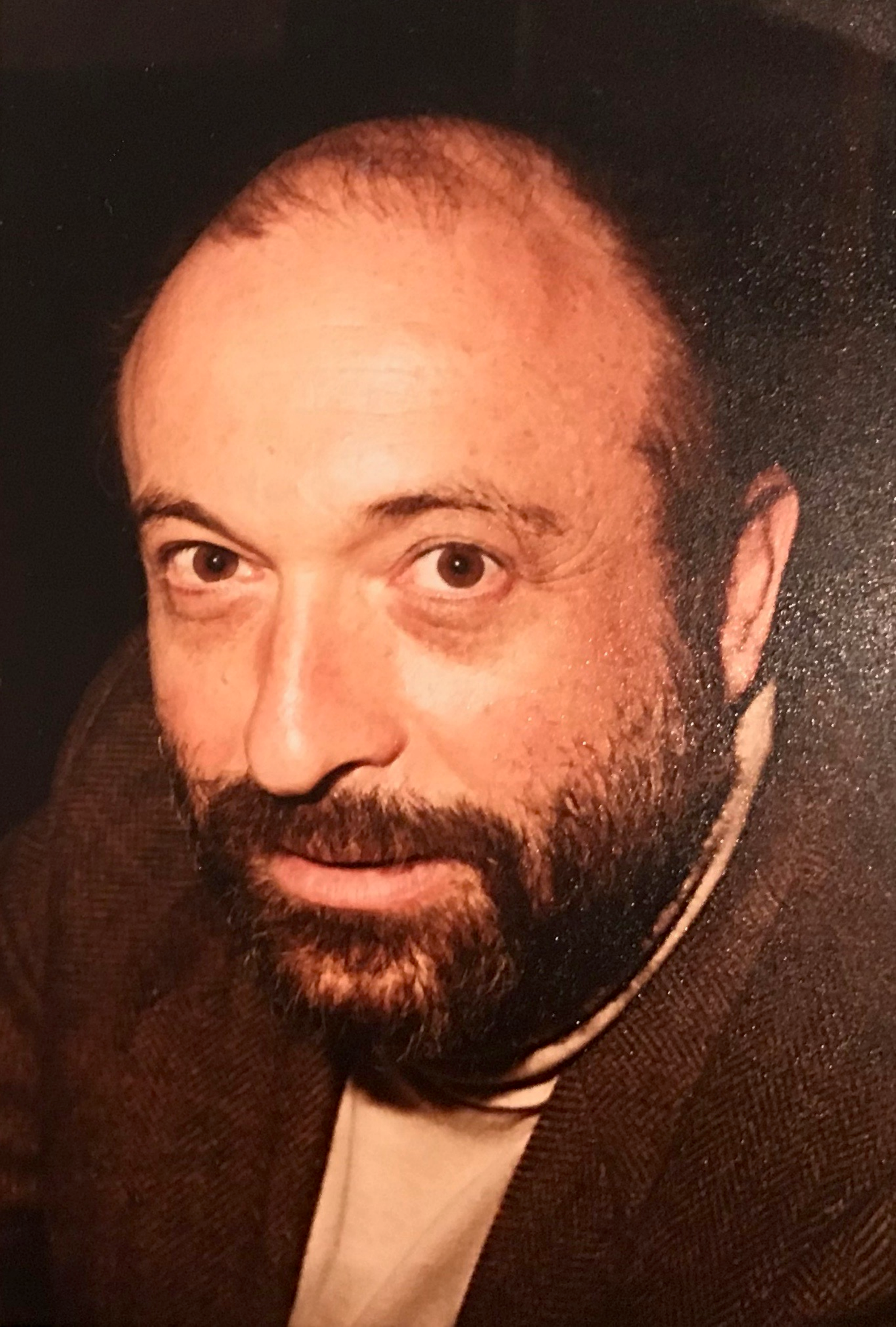}
\caption{Peter G O Freund. 
Credit: The Freund Family}
\label{fig:PGOF}
\end{center}
\end{figure}

We dedicate this special issue to the memory of our esteemed
colleague, inspiring teacher and cherished friend, Peter Freund (figure
\ref{fig:PGOF}).  Peter George Oliver Freund was born on 7 September 1936 in
Timi\c{s}oara, Romania.  He obtained his PhD under the supervision of
Walter Thirring at the University of Vienna in 1960.  He came to the
University of Chicago in 1963, where he remained throughout his
career.  He passed away on 6 March 2018.

Peter's early influential work \cite{Freund:1967hw} proposed
what became known as the Freund-Harari conjecture, whose far-reaching
impact is discussed in the contribution to this special issue by
Veneziano \cite{Veneziano:2020lqs}, see also the review by Peter himself 
\cite{Freund:1900hv}.  Dual resonance models (which were
ultimately understood to be string theories) appeared soon afterwards,
and Peter was among the few theorists at that time who took seriously
the extra dimensions required for their consistency.
Indeed, extra dimensions figured prominently in much of Peter's
subsequent work, including the Cho-Freund paper \cite{Cho:1975sf} that
helped launch a renaissance in Kaluza-Klein theories, the Freund-Rubin
solution \cite{Freund:1980xh} that has played an important role in the
AdS/CFT correspondence, his Kaluza-Klein cosmological solutions
\cite{Freund:1982pg}, and his work \cite{Freund:1984nd} that presaged
the heterotic string.  His book with Applequist and Chodos on
Kaluza-Klein theory \cite{Appelquist:1987} became the standard
reference on the subject.
Many contributions to this special issue are devoted to gravity,
string theory or higher dimensions \cite{Divecchia:2020, Friedan:2017yer, 
Goncharov:2018exa, Hamada:2019fmc, Bergshoeff:2019pij, Henneaux:2019zod, 
Mezincescu:2019vxk, Ketov:2019toi, Sen:2019qit, Fairlie:2020, 
Schwarz:2020emu, Stelle:2020mmg, Bernamonti:2020bcf, Sezgin:2020avr, Das:2020jhy, 
Gunaydin:2019xxl}.

Symmetry was a leitmotif that ran through much of Peter's work.  He
was an early proponent of supersymmetry, which plays a central role 
in string theories that incorporate fermion degrees of freedom..  The Freund-Kaplansky paper
\cite{Freund:1975nr} classified the simple graded Lie algebras; and Peter's book on
supersymmetry \cite{Freund:1986} remains to this day a valuable introduction to
the subject.

Several contributions to this special issue focus on symmetries of 
various kinds \cite{Crampe2019, Kuzenko:2019vvi, Brink:2019zdr, Ananth:2020gkt, Lu:2019rro, 
Harvey:2019qzs, Bars:2020mad, Duff:2020dqb, Ramond:2020dgm}.

In his lectures at UC, Peter highlighted Noether's theorem, which is
also highlighted here in the contribution by Deser \cite{Deser:2019acl}.  But Peter was
also quick to point out that not all conservation laws come from
symmetries.  Indeed, the Arafune-Freund-Goebel paper \cite{Arafune:1974uy} showed
that the 't Hooft magnetic charge is a topological invariant.
Topology also played a key role in the Eguchi-Freund instanton
solution \cite{Eguchi:1976db}.

In later years, Peter became interested in p-adic numbers in physics.
The Freund-Witten paper \cite{Freund:1987ck} on p-adic string
amplitudes and related work \cite{Freund:1987kt, Brekke:1988dg,
Brekke:1993gf} created significant excitement in the theoretical
physics community, briefly recalled here in the contribution by
Frampton \cite{Frampton:2020oeh}.  Although interest in p-adics then
waned, it has recently experienced a renaissance in the context of the
AdS/CFT correspondence, largely due to Steven Gubser, who (together
with his students) contributed to this special issue
\cite{Gubser:2018cha}.  Sadly, this is one of Gubser's final works,
since he died shortly afterwards in a tragic mountaineering accident.

Peter's late work \cite{Freund:1989jq} with his students on exact
S-matrices for integrable perturbations of conformal field theories
was also influential.  Several contributions to this special issue
focus on perturbations of CFTs or integrability \cite{Hoare:2018jim,
Guica:2019vnb, Chakraborty:2019mdf, Barbon:2020amo, Nepomechie:2019tbr}.  The 
contribution by Witten \cite{Witten:2016iux} is a
landmark work on tensor models.

`Multiple dimensions' was both a dominant theme in Peter's work, and
also aptly characterized his other personas.  He was a sonorous baritone, 
who performed in occasional solo recitals.  Schubert was a particular
favorite of his.  He was also a writer, whose works included an
acclaimed book \cite{Freund:2007} on prominent physicists of the 20th
century, as well as a novella and numerous short stories.  He was a
polyglot (including Hungarian, Romanian, German, French and Italian,
besides English) and a raconteur par excellence.  His knowledge of
history and music was encyclopedic; and he had seemingly boundless
energy, invariably brimming with irrepressible excitement over new
ideas.  He was not only a student of Nature's forces, but he was
himself a veritable force of Nature.

Peter is survived by his wife Lucy, his daughters Pauline and
Caroline, and his five grandchildren, for whom he cared deeply and was
very proud.

He is deeply missed by his family, friends, colleagues and students.

\section*{Acknowledgments}

We are grateful to the many friends and colleagues of Peter who
contributed to this special issue.  JH acknowledges support from the
NSF under grant PHY 1520748; EM acknowledges support from the DOE
under grant DE-SC0009924; and RN acknowledges support from a Cooper
fellowship.

\providecommand{\href}[2]{#2}\begingroup\raggedright\endgroup

\end{document}